\documentclass{jfm}
\usepackage{graphicx}
\usepackage{newtxtext}
\usepackage{newtxmath}
\usepackage{natbib}
\usepackage{hyperref}
\usepackage{bm}
\usepackage{caption}
\usepackage[normalem]{ulem}
\usepackage[dvipsnames]{xcolor}
\usepackage{lineno}
\usepackage{pdfpages}
\hypersetup{
    colorlinks = true,
    urlcolor   = blue,
    citecolor  = CadetBlue,
}

\title{Thermal convection in one, two, three and four dimensions}

\author{Ambrish Pandey\aff{1,2},
        \corresp{\email{ambrish.pandey@ph.iitr.ac.in}}
        Harshit Tiwari\aff{3},
        \and Katepalli R. Sreenivasan\aff{2,4}
        }

\affiliation{\aff{1} Department of Physics, Indian Institute of Technology Roorkee, Roorkee 247667, Uttarakhand, India
\aff{2} Center for Astrophysics and Space Science, New York University Abu Dhabi, Abu Dhabi 129188, United Arab Emirates \\
\aff{3} Department of Physics, Indian Institute of Technology Kanpur, Kanpur 208016, India \\
\aff{4} Tandon School of Engineering, Department of Physics and Courant Institute of Mathematical Sciences, New York University, New York, NY 11201, USA
}

\def\au#1{#1} \def\yr#1{#1} \def\at#1{#1} \def\jt#1{\textit{#1}}
\def\bt#1{#1} \def\bvol#1{\textbf{#1}} \def\vol#1{#1} \def\pg#1{#1}
\def\publ#1{#1} \def\arxiv#1{#1} \def\org#1{#1} \def\st#1{\textit{#1}}

\begin{document}
\maketitle


\begin{abstract}
We study by means of direct numerical simulations the influence of the dimensionality of convection on flow properties. We call attention to a few general principles from considering in totality the results from one dimension, two dimensions, three dimensions and four dimensions. In particular, we explore two practical aspects: (1)  the transient time, or the amount of time it takes for the flow to reach the steady state; and (2) possible implications for the so-called ultimate state.  
\end{abstract}



\section{Introduction}
\label{sec:intro}

An increasing fraction of basic research on turbulent thermal convection is being drawn in recent years from direct numerical simulations (DNS). With this development, it has also become clearer that more attention needs to be paid to numerical convergence, resolution, transient time to steady state, etc. In this paper we attempt to understand transient times needed to achieve the steady state of convection in 1D, 2D, 3D and 4D, and infer the advantages (or lack thereof) of studying convection in these different dimensions.

\citet{Pandey:JFM2025}---PS henceforth---obtained transient times from the DNS of 2D Rayleigh-B\'enard convection (RBC) in a box of aspect ratio $\Gamma = 1$, and Prandtl number $Pr$ = 0.1 and 1, for Rayleigh number $Ra$ between $10^6$ and $10^{12}$. (We define $Ra$, $Pr$ and $\Gamma$ in \S 2.) A main conclusion was that long transients with large scale fluctuating heat transport accompany 2D convection, which made observations of the ultimate state~\citep{Kraichnan:POF1962} quite uncertain. By the ultimate state, we refer to the asymptotic high-$Ra$ regime where the Nusselt number scaling surpasses the classical $1/3$ exponent and approaches a power of $1/2$ (with possible logarithmic corrections).

As another variant of the practically important problem of 3D convection, one may consider the case of 4D, for which the flow occurs in four spatial dimensions and time.
There is no instance in the universe for which this configuration applies, yet there are good reasons to consider it. For instance, the Ising universality class has its upper critical dimension equal to 4. That is, for $D \ge 4$ the renormalization fixed point controlling the transition is Gaussian, so the leading critical exponents assume mean-field values, possibly with logarithmic corrections. It has been a long-held speculation that turbulence in 4D may also have normal scaling with no anomalies. For instance, will the 4D convection adopt to the ultimate state at a manageably smaller Rayleigh number than in 3D?

To obtain a more complete perspective, we also consider heat transport in the 1D setting. A moment’s thought shows that, when the bottom of a 1D convection system is heated and its top cooled, heat can be conveyed up only by conduction; there can never be any convection at any $Ra$ or $Pr$. (One may interpret this result to mean that the transient state to convection is infinitely long.) In compressible cases, pressure perturbations can travel up and down but, again, there can be no convection. The oscillatory character of pressure has a dependence on $Pr$ but the basic result is the same. We present the results only briefly in the body of the text and relegate details to the Supplementary Material (SM).

This paper is thus a study of the influence of dimensionality on thermal convection. \S\ref{sec:sim_details} gives a brief discussion of the simulation parameters, while \S\ref{sec:results} discusses the main results. It concludes with a few summary remarks and outlook in \S\ref{sec:summary}. To be concise, we consistently refer to RBC in {horizontally-}periodic boxes as RBC-P. The numerical code {\sc Dhara} used to compute RBC-P is described and validated in SM.

\section{Simulation details}
\label{sec:sim_details}

We perform DNS in 2D, 3D and 4D settings assuming the Oberbeck-Boussinesq model
\begin{eqnarray}
\nabla \cdot {\bm u} & = & 0, \label{eq:m} \\ 
\frac{\partial {\bm u}}{\partial t} + {\bm u} \cdot \nabla {\bm u} & = & - \frac{ \nabla p}{\rho_0} + \alpha g (T-T_0) \hat{z} + \nu \nabla^2 {\bm u}, \label{eq:u} \\ 
\frac{\partial T}{\partial t} + {\bm u} \cdot \nabla T & = & \kappa \nabla^2 T, \label{eq:T}
\end{eqnarray}
where ${\bm u}({\bm r})$, $p({\bm r})$, and $T({\bm r})$ are, respectively, velocity, pressure, and temperature. The space dimensionality of the system is to be treated as appropriate. The acceleration due to gravity is $-g \hat{z}$; $\rho_0$ is the reference density, $T_0$ is the reference temperature, and $\alpha, \nu, \kappa$ are the isobaric thermal expansion coefficient, kinematic viscosity, and thermal diffusivity, respectively. The dynamics of RBC is governed by $Ra = \alpha g \Delta T H^3 / (\nu \kappa)$ and $Pr = \nu / \kappa$, where $\Delta T$ is the applied temperature difference between the bottom and top surfaces of the fluid layer of depth $H$. The flow velocities are measured in units of the free-fall velocity $u_f = \sqrt{\alpha g \Delta T H}$ and time scale in units of free-fall time $t_f = H/u_f$. 

\subsection{DNS of 2D convection}
In 2D convection, the flow is constrained to a vertical plane, with ${\bm u}({\bm r}) = (u_x, u_z)$ with ${\bm r} = (x, z)$. We first study the case of no-slip condition at all walls, the horizontal walls being isothermal and the vertical walls adiabatic. In PS, we have already explored the characteristics of transients for $Pr = 0.1$ and $Pr = 1$ and for $10^6 \le Ra \le 10^{12}$. Here, we include $Pr$ of $0.05$ and $0.021$. DNS of 2D RBC, governed by equations~\eqref{eq:m}--\eqref{eq:T}, are conducted using a spectral element solver {\sc Nek5000}~\citep{Fischer:JCP1997}. The square domain is decomposed into $N_e$ elements, further discretized using $N^\mathrm{th}$-order Legendre polynomials. Details of DNS for $Pr = 0.1$ and $Pr = 1$ can be found in PS, and those for $Pr = 0.021$ in \citet{Pandey:JFM2021}. Additional details for $Pr = 0.021$ and $Pr = 0.05$ are provided in table~\ref{table:sim_detail_2d}.

\begin{table}
\captionsetup{width=1\textwidth}
  \begin{center}
\def~{\hphantom{0}}
  \begin{tabular}{lcccccc}
$Pr$ & $Ra$ & $N_e N^2$ & $Nu$ & $Re$ & $t_{sim} \, (t_f)$ 
\vspace{2mm} 
\\
0.021 &         $10^{7}$ &  $1150^2$ 	 & 	 10.97 $\pm$    4.6 &  22570 $\pm$   4986 &    394 \\ 
0.021 & $2\times 10^{7}$ &   $900^2$ 	 & 	 13.04 $\pm$    5.2 &  33921 $\pm$   7101 &    557 \\ 
0.021 & $3\times 10^{7}$ &  $1150^2$ 	 & 	 14.42 $\pm$    5.7 &  43040 $\pm$   8712 &    496 \\ 
0.021 & $5\times 10^{7}$ &  $2070^2$ 	 & 	 16.58 $\pm$    7.0 &  59928 $\pm$  11542 &    136 \\ 
0.021 &         $10^{8}$ &  $2530^2$ 	 & 	 19.68 $\pm$     10 &  97012 $\pm$  17184 &    220 \\ 
0.021 & $4\times 10^{8}$ &  $3122^2$ 	 & 	 28.32 $\pm$     20 & 275539 $\pm$  19188 &    106 \\ 
0.021 &         $10^{9}$ &  $4014^2$ 	 & 	 37.43 $\pm$     28 & 443526 $\pm$  53505 &    159 \\ 
0.05 &         $10^{7}$ &   $700^2$ 	 & 	 12.41 $\pm$    5.1 &  11624 $\pm$   2588 &   1177 \\ 
0.05 & $2\times 10^{7}$ &   $700^2$ 	 & 	 14.52 $\pm$    6.8 &  16135 $\pm$   3401 &   1370 \\ 
0.05 & $3\times 10^{7}$ &   $900^2$ 	 & 	 16.06 $\pm$    7.6 &  20167 $\pm$   4497 &   1008 \\ 
0.05 &         $10^{8}$ &  $1150^2$ 	 & 	 21.64 $\pm$     10 &  41298 $\pm$   7966 &    763 \\ 
0.05 & $3\times 10^{8}$ &  $1610^2$ 	 & 	 29.27 $\pm$     16 &  82321 $\pm$  20101 &    632 \\ 
0.05 & $5\times 10^{8}$ &  $1610^2$ 	 & 	 33.50 $\pm$     20 & 122255 $\pm$  25627 &    882 \\ 
0.05 &         $10^{9}$ &  $2478^2$ 	 & 	 41.51 $\pm$     27 & 185128 $\pm$  30221 &    250 \\
  \end{tabular}
  \caption{Important DNS parameters of RBC in a 2D box of $\Gamma = 1$: {$N_e N^2$ is the total number of mesh cells; $t_{sim}$ is the total duration of simulation in the nominally steady state; $Nu$ and $Re$ are, respectively, the global heat and momentum transports in this state; the error bars are the corresponding standard deviations.}} 
  \label{table:sim_detail_2d}
  \end{center}
\end{table}

To make direct comparisons with $D=3$ and $D=4$, we have performed simulations of 2D convection also {for $\Gamma = 1$}, with periodic sidewalls (RBC-P). These simulations are conducted using the incompressible module of the finite-difference code {\sc Dhara} (see SM). The gross flow response quantities are summarized in table~\ref{table:dim_comparison_complete}.

\begin{table}
\captionsetup{width=1\textwidth}
\begin{center}
\def~{\hphantom{0}}
\begin{tabular}{lc|ccc|ccc|ccc}

\vspace{1mm} 

& & \multicolumn{3}{c|}{$D=2$ } & \multicolumn{3}{c|}{$D=3$ } & \multicolumn{3}{c}{$D=4$ } \\

\vspace{2mm} 

$Ra$ & $N$ & $Nu$ & $Re$ & $t_{sim}$ & $Nu$ & $Re$ & $t_{sim}$ & $Nu$ & $Re$ & $t_{sim} $ \\ 
$2 \times 10^5$ & $64$  & $4.8 \pm 3.8$ & $94 \pm 30$ & 200 & $7.4 \pm 0.1$ & $98 \pm 1$ & 200 & $7.9 \pm 0.1$ & $100 \pm 1$ & 200\\
$5 \times 10^5$ & $64$  & $6.2 \pm 3.2$ & $173 \pm 19$ & 500 & $8.9 \pm 0.4$ & $145 \pm 6$ & 200 & $9.2 \pm 0.3$ & $147 \pm 6$ & 200\\
$10^6$          & $64$  & $6.9 \pm 5.3$ & $254 \pm 34$ & 500 & $10.4 \pm 0.8$ & $202 \pm 6$ & 200 & $11.1 \pm 0.3$ & $205 \pm 6$ & 200\\
$2 \times 10^6$ & $64$  & $8.5 \pm 5.2$ & $385 \pm 33$ & 500 & $11.3 \pm 0.7$ & $277 \pm 7$ & 200 & $13.1 \pm 0.3$ & $278 \pm 7$ & 200\\
$5 \times 10^6$ & $128$ & $10.3 \pm 5.4$ & $656 \pm 29$ & 500 & $14.3 \pm 0.6$ & $412 \pm 20$ & 200 & $16.1 \pm 0.5$ & $411 \pm 20$ & 200\\
$10^7$          & $128$ & $11.8 \pm 4.2$ & $955 \pm 19$ & 500 & $17.5 \pm 0.8$ & $592 \pm 13$ & 200 & $19.9 \pm 0.4$ & $577 \pm 13$ & 200\\
$2 \times 10^7$ & $128$ & $13.1 \pm 4.5$ & $1457 \pm 22$ & 500 & $20.6 \pm 0.8$ & $780 \pm 18$ & 200 & $23.7 \pm 0.6$ & $793 \pm 18$ & 200\\
$5 \times 10^7$ & $256$ & $13.3 \pm 5.5$ & $2595 \pm 21$ & 1000 & $26.5 \pm 0.8$ & $1229 \pm 28$ & 200 & $31.1 \pm 0.7$ & $1230 \pm 28$ & 100\\
$10^8$          & $256$ & $13.6 \pm 5.5$ & $3865 \pm 20$ & 1000 & $32.6 \pm 0.8$ & $1624 \pm 30$ & 200 & $37.6 \pm 0.8$ & $1680 \pm 30$ & 100\\
$2 \times 10^8$ & $256$ & $15.5 \pm 6.9$ & $5696 \pm 22$ & 1000 & $39.1 \pm 0.8$ & $2333 \pm 42$ & 200 & $45.7 \pm 0.9$ & $2390 \pm 42$ & 100\\
$5 \times 10^8$ & $512$ & $20.4 \pm 13.4$ & $9933 \pm 30$ & 1000 & --- & --- & --- & --- & --- & --- \\
$10^9$          & $512$ & $24.6 \pm 18.1$ & $14624 \pm 38$ & 1000 & --- & --- & --- & --- & --- & --- \\
\end{tabular}
\caption{Response quantities for RBC-P in $D=2, 3,$ and $4$, $Pr=1$. {$N$ is the number of grid points for each spatial dimension (accounting for a total of $N^D$ points)} used by the finite-difference solver {\sc Dhara}. $Nu$ and $Re$ values are time-averaged, with $\pm$ one standard deviation; $t_{sim}$ is the integration time in units of $t_f$.}
\label{table:dim_comparison_complete}
\end{center}
\end{table}

\subsection{DNS of 3D convection}

We consider the following configurations {with no-slip and isothermal top and bottom walls: (a) closed cube with no-slip condition at all the walls}; (b) closed rectangular cuboid of length $2.4H$ and width $0.8H$~\citep{Pandey:JFM2026}, especially to understand the effects of geometry; and (c) cube with periodic sidewalls (RBC-P). {The sidewalls in (a) and (b) are adiabatic.} The DNS of RBC are performed for $Pr = 0.1, 1, 10$ for $Ra$ between $10^6$ and $10^{10}$. The higher-$Ra$ simulations are carried out on coarser meshes than demanded for a faithful replication of the entire flow structure, but the conclusions on transients do not depend on the resolution, up to a point. {In SM, we examine the consequences of using coarser resolutions on transients; see also PS.} The RBC-P cases for $Pr = 1$ are conducted using the incompressible module of the finite-difference code {\sc Dhara}, as summarized in table~\ref{table:dim_comparison_complete}.

\subsection{DNS of 4D convection}

4D convection is performed in a periodic hypercube of size $L_w = L_x = L_y = H$ for $Pr = 1$ and Rayleigh numbers between $Ra = 10^5$ and $2 \times 10^8$. To ensure a direct comparison, 2D and 3D simulations are conducted using a consistent numerical framework, as summarized in table~\ref{table:dim_comparison_complete}. These cases utilize the incompressible module of the finite-difference solver {\sc Dhara}, which employs a second-order spatial discretization on a staggered marker-and-cell (MAC) grid and a third-order Runge-Kutta scheme for time integration.

The computational complexity of 4D simulations increases strongly with $Ra$. Even for $Ra = 2 \times 10^8$, which was resolved on a $256^4$ grid, it represented approximately 4.3 billion degrees of freedom per variable. This specific run was executed on 128 NVIDIA A100 GPUs across 32 compute nodes on the Polaris supercomputer at the Argonne Leadership Computing Facility (ALCF). This simulation required approximately 18 wall-clock hours to reach an integration time of $t_{sim} = 100$. 

\section{Results}
\label{sec:results}

\subsection{Basic response quantities}
The instantaneous state of convection is given by the domain-averaged kinetic energy
\begin{equation}
E  = \langle \vert {\bm u} \vert^2 /2 \rangle_V / u_f^2 \, .
\end{equation}
$E = 0$ in the purely diffusive state of no motion, and $E > 0$ represents a perturbed state. Temporal evolution of the flow can be described by the evolution of $E$. Once convection is established and the system is in a statistically steady state, the flow strength is measured by $Re$, which is commonly defined in DNS studies using the root-mean-square (RMS) velocity
\begin{equation}
    Re = u_{RMS} H / \nu , \quad \mbox{where} \quad u_{RMS} = \sqrt{ \langle \vert {\bm u} \vert^2 \rangle_{V,t} } \, .
\end{equation}
Here $\langle \cdot \rangle_{V,t}$ is the average over the entire domain, and also time in the statistically steady state. The mean kinetic energy in this state is defined as $E_{av} = 0.5 u_{RMS}^2 / u_f^2$.

The heat transport through the convective layer is quantified using the Nusselt number
\begin{equation}
    Nu = 1 + \frac{H}{\kappa \Delta T} \langle u_z T \rangle_{V,t} \, . \label{eq:Nu}
\end{equation}
The globally averaged kinetic energy and thermal dissipation rates can also be used to define Nusselt numbers; their closeness to $Nu$ from \eqref{eq:Nu} is thought to ensure the spatial and temporal convergence (see \citet{Stevens:JFM2010} and PS). We have verified that the Nusselt numbers computed using different methods agree within a few percent.

\subsection{1D flow}

Conduction is the only means of transport between top and bottom walls in the incompressible case. 
While the continuity equation, $\partial (\rho u_z) / \partial z = -\partial \rho / \partial t$ {(where $\rho$ is the density of the fluid)}, allows for local density variations in the compressible case, the strictly one-dimensional geometry and impenetrable boundaries at the top and bottom prevent vertical mass flux or bodily overturning---hence no convection. A linear instability analysis and the DNS (both detailed in SM) confirm this constraint. The determinant of the linearized system is strictly non-zero for all real wavenumbers, mathematically precluding the stationary bifurcation required for classical convection. The system does support oscillatory acoustic modes as $Ra$ increases, but these purely longitudinal compressions do not lead to heat transport with a unidirectional motion. (Professor John Wettlaufer has suggested a possible connection of the present 1D considerations to highly confined quasi-1D superfluid systems: there will be no convection in such systems either, but its detailed discussion here takes us too far from the theme of the paper.)

\subsection{2D convection}

\begin{figure}
\captionsetup{width=1\textwidth}
\centering
\includegraphics[width=0.95\textwidth]{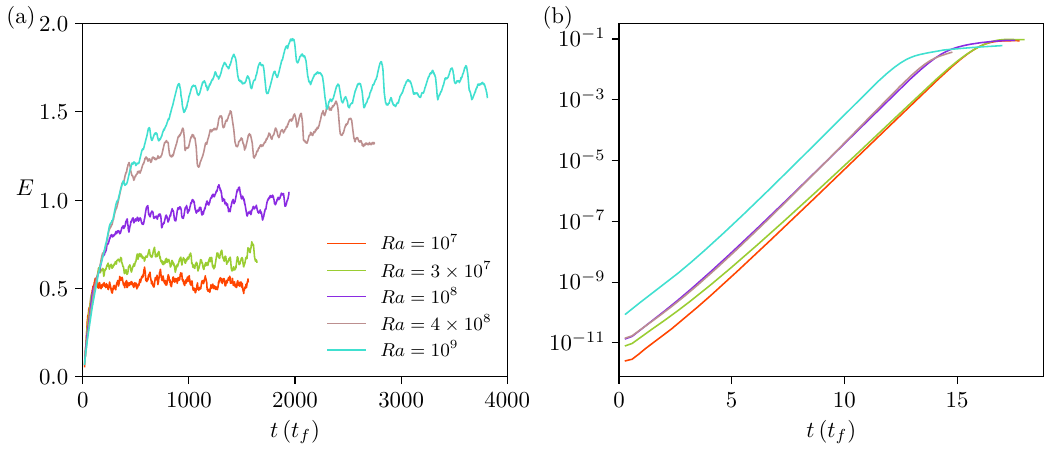}
\caption{(a) Evolution of $E$ for different Rayleigh numbers at $Pr = 0.021$ in a closed 2D square box. Time required to achieve a nominally steady state increases with $Ra$. (b) Energy growth is exponential during the initial phase when the perturbations grow and convective motion is being established. (There are on the order of 100 data points in the initial exponential part.)}
\label{fig:E_t_2d}
\end{figure}

We showed in PS, for $10^7 \le Ra \le 10^{12}$, that the scaling of the transient time varies in the range $Ra^{0.6}$ to $Ra^{0.7}$, consistent with Lindborg's theory \citep{Lindborg:2025}, but the coefficient of proportionality depended on $Pr$. We note that Lindborg's theoretical lower bound corresponds to the global diffusion time $H^2/\nu$, and the transient times observed in our simulations actually exceed this timescale. Recall that the RMS velocity of turbulent convective flows in a 2D domain does not scale as the free-fall velocity; it is a function of $Ra$ and $Pr$. PS found that $u_{RMS} / u_f$ scales as $Ra^{1/6} Pr^{-1/2}$ in high-$Re$ regime. This scaling suggests that, in 2D RBC, it takes longer with increasing $Ra$ or decreasing $Pr$ to achieve the nominally steady state. This is demonstrated in figure~\ref{fig:E_t_2d}, where temporal evolutions of $E$ in a 2D square box are shown for various $Ra$, and $Pr = 0.021$. Figure~\ref{fig:E_t_2d}(b) shows that $E$ in the initial phase grows exponentially when the convective motion is getting established. However, as seen better in figure~\ref{fig:E_t_2d}(a), this initial exponential is followed by another slow exponential growth in the convection state. Only after this extended second growth state does $E$ begin to attain a nominal mean that increases with $Ra$, with huge fluctuations around it.

For a simulation performed {\it ab initio} from the conduction state, $E$ evolves in the slow exponential state of convection according to
\begin{equation}
    [E_{av} - E(t)]/E_{av} = c \exp(-kt) \, , \label{eq:E_t_trns}
\end{equation}
where $c$ is the scaling factor and $k$ is the growth rate (see PS). The transient time $t_{trns}$ is defined as the time that is required for the flow to reach a state, where $E$ is a certain fraction of $E_{av}$. Taking, for example, $E = 0.96 E_{av}$ in \eqref{eq:E_t_trns}, the transient time is computed as
\begin{equation}
    t_{trns} = \frac{1}{k} \log \frac{c}{0.04} \, . \label{eq:t_trns}
\end{equation}

\begin{figure}
\captionsetup{width=1\textwidth}
\centering
\includegraphics[width=1\textwidth]{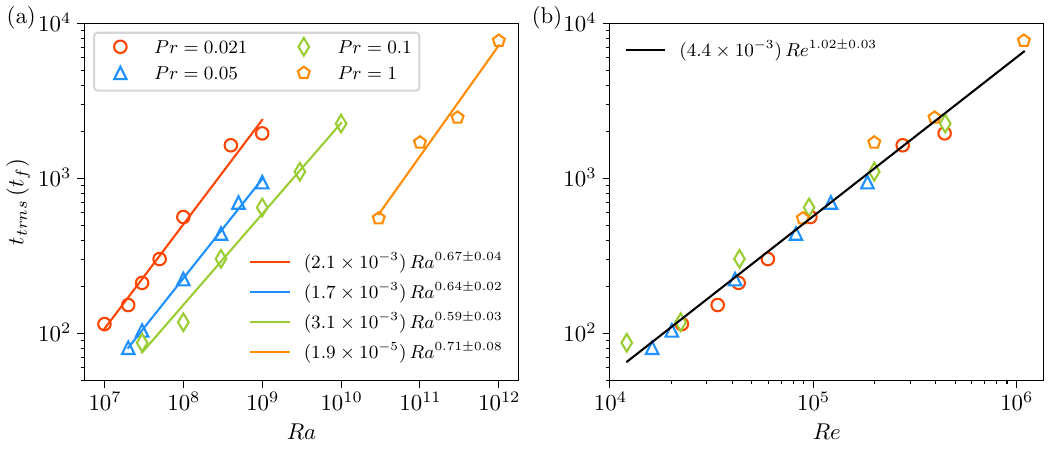}
\caption{Transient time $t_{trns}$ {for RBC in a 2D square box} as a function of (a) Rayleigh number and (b) Reynolds number. Solid lines represent best fits. Scaling exponents in (a) depend on $Pr$. {Solid black line in (b) is the best fit using data for all $Pr$.} Data for $Pr = 1$ and $Pr = 0.1$ are taken from PS.}
\label{fig:t_trns_2D}
\end{figure}

Transient times $t_{trns}$ estimated using \eqref{eq:E_t_trns}, plotted against $Ra$ in figure~\ref{fig:t_trns_2D}(a), show an increase with $Ra$. The best fit exponents vary between 0.59 and 0.71 depending on $Pr$. Consistent with the trend found in PS, they become longer with decreasing $Pr$ for a fixed thermal forcing. Figure~\ref{fig:t_trns_2D}(b) shows $t_{trns}$ as a function of the flow Reynolds number; data for different Prandtl numbers nearly collapse and the best fits are approximately linear in $Re$. The best fit for data at all $Pr$ yields $t_{trns} = 0.0044 Re^{1.02}$.

It should be stressed that these data are applicable for \textit{ab initio} calculations (i.e., convection begun from slightly perturbed conduction states). As the steady state kinetic energy is higher for higher $Ra$, there will be a transient period also when the simulation is begun from the lower $Ra$. However, those transient times will be shorter than for the \textit{ab initio} simulations. The observed linear dependence of the transient time on $Re$ imposes severe restrictions on the exploration of 2D convection at high $Re$—as could, in fact, be inferred already from \citet{Lindborg:2025}.

\subsection{3D convection}

\begin{figure}
\captionsetup{width=1\textwidth}
\centering
\includegraphics[width=1\textwidth]{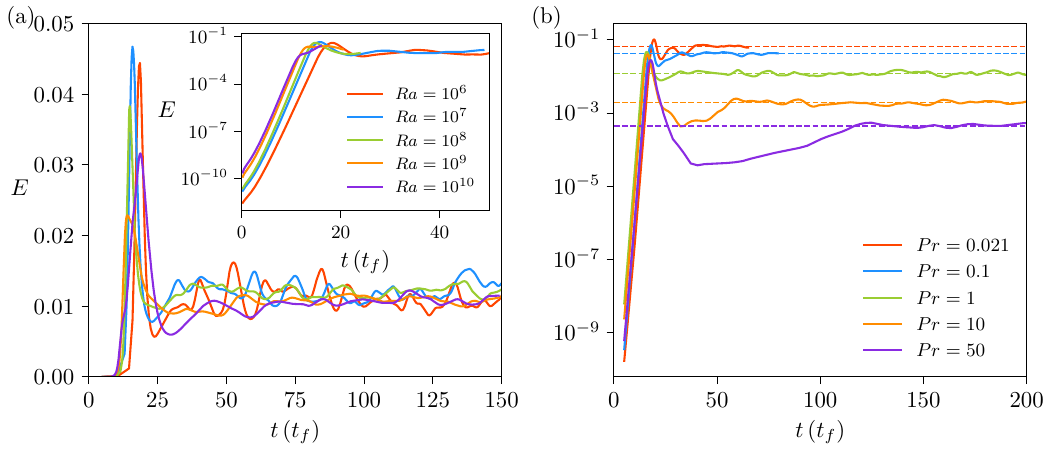}
\caption{Temporal evolution of $E$ for RBC in a closed cubic domain (a) for various $Ra$ at $Pr = 1$, and (b) for different flows at $Ra = 10^7$. Inset in (a) reveals that the growth in the initial phase during the setting up of the convective motion is exponential in nature. Dashed horizontal lines in (b) indicate $E_{av}$.} 
\label{fig:E_t_3d}
\end{figure}

This is the practically relevant case. In figure~\ref{fig:E_t_3d}(a), the evolutions of $E$ {for RBC} in a closed cubic domain for $Pr = 1$ and $10^6 \le Ra \le 10^{10}$ are shown, where the simulations begin from the diffusive state. We observe that $E$ grows rapidly in the beginning and starts to fluctuate, achieving a plausible $E_{av}$ beyond $50$ to $70 \, t_f$. The inset of figure~\ref{fig:E_t_3d}(a), where $E$ is shown on a logarithmic scale, reveals that the initial growth is exponential {as in 2D}, when the convective motion is getting established. Qualitatively similar evolutions of $E$ occur for flows at $Pr = 0.1$ and $Pr = 10$. The same is true also for flows in a rectangular cuboid of dimensions $L_x = 2.4H$ and $L_y = 0.8H$~\citep{Pandey:JFM2026}{---see SM}. Note that, unlike in 2D, the variation of $E_{av}$ with $Ra$ is weaker for $Pr \approx 1$, though $E_{av}$ decreases slightly with $Ra$ for low-$Pr$ fluids, while it increases slightly for high-$Pr$ fluids. This is also inferred from the scaling of the Reynolds number in 3D convection: $Re \sim Ra^{1/2}$ for $Pr \sim 1$~\citep{Chilla:EPJE2012, Pandey:JFM2026}, $Re \sim Ra^{0.44}$ for $Pr \lesssim 0.03$~\citep{Pandey:PoF2016, Scheel:PRF2017}, and $Re \sim Ra^{0.6}$ for $Pr \gg 1$~\citep{Silano:JFM2010, Horn:JFM2013, Pandey:PRE2014}. 

Evolution of $E$ for different $Pr$ and $Ra = 10^7$ in the same cubic domain is shown in figure~\ref{fig:E_t_3d}(b). Note that $E_{av}$, and thus $Re$, decreases with increasing $Pr$~\citep{Pandey:EPL2021}. The logarithmic scale for $E$ in figure~\ref{fig:E_t_3d}(b) highlights the exponential growth in the beginning, and also a strong dependence of $E_{av}$ on $Pr$. The steady state takes slightly longer with increasing $Pr$; thus, it seems that the transient time in 3D increases, albeit weakly, with decreasing $Re$. However, as a whole, the trends with respect to $Ra$ or $Re$ are not strong as in 2D (as one might indeed have expected).

\subsection{4D convection}

\subsubsection{Motivation}
Here, we consider convection in four spatial dimensions and time. Since 4D convection never appears in the universe, a few sentences are useful to justify the effort. We recall that~\citet{Wilson:PRL1972} considered systems with spatial dimension of $4 - \epsilon$, where $\epsilon << 1$. Their insight was that 4D is the upper critical dimension for equilibrium phase transitions, and that Landau’s mean-field theory becomes exact for 4D and beyond, with deviations from it arising in fewer dimensions. Indeed, they showed that one can treat the deviation perturbatively in $\epsilon$.
\citet{Nelkin:PRA1974, Nelkin:PRA1975} proposed that four spatial dimensions might play an analogous role in turbulence theory, and argued that intermittency arises from fluctuations of the local energy dissipation and cascade transfer rates, which would weaken with increasing dimension. 
If this conjecture is true, $D = 4$ would behave like a mean-field limit for turbulence. This has never been proved despite some suggestion~\citep{Gotoh:PRE2007, Yamamoto:PRE2012} that intermittency weakens with increasing dimension. The last two papers studied homogeneous and isotropic turbulence. 

One of the problems in drawing firm conclusions from these studies is that the DNS in higher dimensions becomes more expensive and so the Reynolds numbers of the DNS are smaller. Although, because of this, it was clear from the outset that the Rayleigh numbers achievable would be quite modest in 4D, we regarded that some instructive lesson could arise while making comparisons with 3D. 

\subsubsection{Results}

The results of 4D simulations are summarized in table~\ref{table:dim_comparison_complete} for $Pr = 1$. Due to the high computational cost associated with the $N^4$ grid scaling, these runs currently span Rayleigh numbers from $10^5$ to $2 \times 10^8$, but are adequate for drawing a preliminary conclusion. 

\begin{figure}
\captionsetup{width=1\textwidth}
\centering
\includegraphics[width=0.55\textwidth]{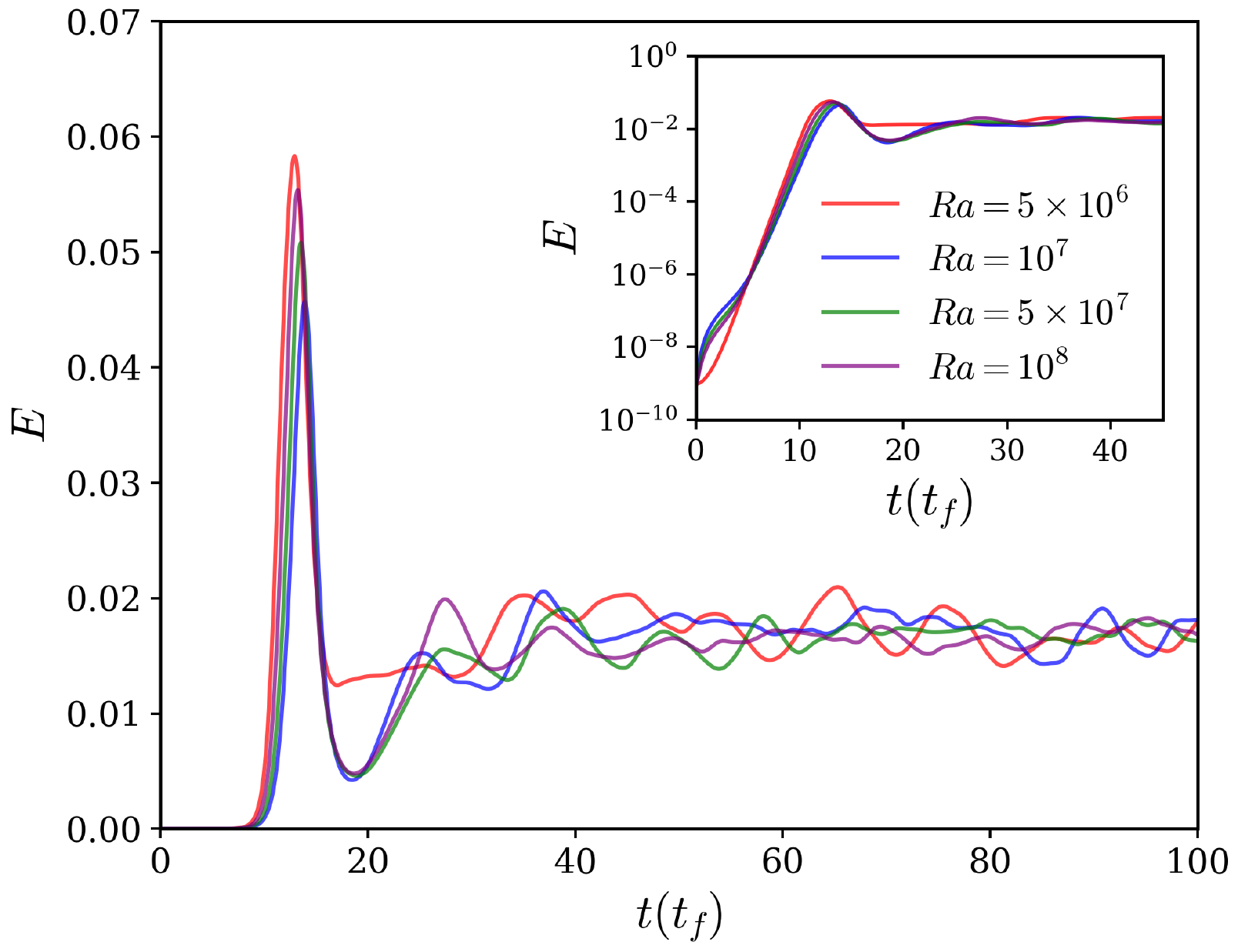}
\caption{Temporal evolution of the domain-averaged kinetic energy $E$ in a 4D horizontally-periodic domain for $Pr = 1$ at various Rayleigh numbers. Similar to the 3D case, the inset reveals an initial phase of exponential growth as the convective motion is established, followed by a statistically steady state where $E$ fluctuates around its mean value $E_{av}$.}
\label{fig:E_t_4d}
\end{figure}

As shown in figure~\ref{fig:E_t_4d}, the temporal evolution of the kinetic energy $E$ in 4D is similar to that in 3D. After a slightly perturbed conduction state, there is an initial exponential growth phase in which the convective motion is established. The system later enters a statistically steady state characterized by fluctuations around the mean value $E_{av}$. The transient time for reaching this steady state in 4D is approximately $40$ to $50 \, t_f$, which is somewhat smaller than in 3D but comparable overall. The transient times show an increasingly weaker trend with $Re$ as the dimensionality increases beyond 2; the 3D and 4D cases are similar in this regard.

\begin{figure}
\captionsetup{width=1\textwidth}
\centering
\includegraphics[width=1\textwidth]{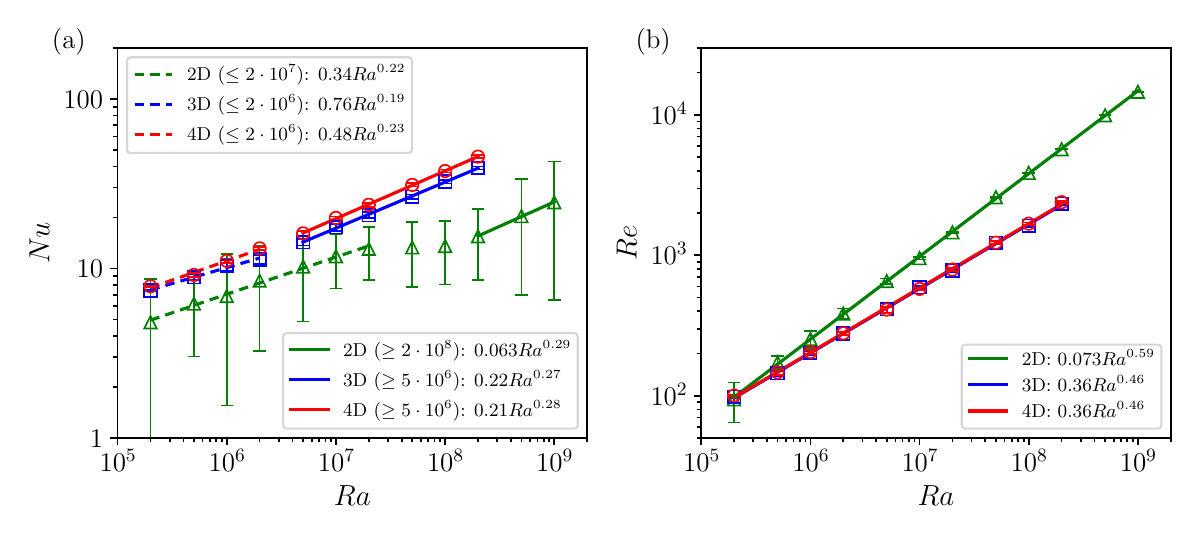}
\caption{Scaling of global response quantities with $Ra$ for RBC-P at $Pr = 1$ in $D=2, 3,$ and $4$. (a) Nusselt number ($Nu$) versus $Ra$. Dashed and solid lines represent power-law fits for low- and moderate-$Ra$ regimes, respectively. (b) Reynolds number ($Re$) versus $Ra$ with global power-law fits. Error bars denote the standard deviation of temporal fluctuations.}
\label{fig:Nu_Re_scaling}
\end{figure}

However, the Nusselt and Reynolds numbers, presented in figure~\ref{fig:Nu_Re_scaling}, depend on the dimensionality of convection. Figure~\ref{fig:Nu_Re_scaling}(a) shows that the Nusselt number at a given $Ra$ increases with $D$, though the difference is less conspicuous between 3D and 4D than it is between 2D and 3D. For example, in the range $Ra \geq 5 \times 10^6$, the scalings for 3D ($Nu = 0.22 Ra^{0.27}$) and 4D ($Nu = 0.21 Ra^{0.28}$) share closely similar powers and prefactors. By comparison, the scaling in 2D follows $Nu = 0.063 Ra^{0.29}$, though characterized by massive temporal fluctuations. The overall enhancement of $Nu$ in higher dimensions stems from the increased spatial degrees of freedom, which allow thermal plumes to advect heat with significantly fewer disruptive interactions. For $Ra = 2 \times 10^8$, $Nu$ in 3D is roughly $2.4$ times that in 2D, while that in 4D it is only 15\% higher than that in 3D. 2D data show a conspicuous break for a decade around $Ra = 10^8$, which is due to the appearance of a strong zonal flow~\citep{Goluskin:JFM2014} for this range of $Ra$. The flow structures and their evolutions suggest that 2D convective flows in a horizontally-periodic box of $\Gamma = 1$ comprise a strong zonal component for all higher-$Ra$ cases in the present work. Consequently, the heat transport is lower than that in a closed square box, as reported by \citet{Pandey:JFM2025}.

Figure~\ref{fig:Nu_Re_scaling}(b) further highlights a striking difference between 2D, on the one hand, and 3D and 4D, on the other: the Reynolds number is higher in 2D ($Re = 0.071 Ra^{0.59}$) compared to the lower (and nearly indistinguishable) scalings of 3D and 4D ($Re \approx 0.36 Ra^{0.46}$). Rather than indicating more intense small-scale turbulence, this elevated $Re$ in 2D is a consequence of enstrophy conservation, which prohibits vortex stretching and forces an inverse cascade of kinetic energy. The resulting accumulation of energy in the largest available scales generates a domain-filling large-scale circulation that inflates the RMS velocity. In 3D and 4D, the availability of vortex stretching enables a forward energy cascade to dissipative scales, severely weakening this domain-sized scale and yielding lower Reynolds numbers. Note that the scaling exponent in 2D in the horizontally-periodic domain is smaller than $2/3$, which was derived by \citet{Lindborg:2025} and confirmed numerically by \citet{Pandey:JFM2025}. However, we stress that the scaling $Re \sim Ra^{2/3}$ is realized only at high enough Rayleigh numbers. \cite{Pandey:JFM2025}  observed that the $Re$ in the closed square domain exhibits a shallower scaling at lower Rayleigh numbers. In fact, their data for $Ra$ between $10^7$ and $10^9$ for $Pr = 1$ is consistent with a $Re \sim Ra^{0.59}$ scaling, very similar to that found here.

These scaling behaviors also provide a context for commenting on convection at extreme thermal forcing. One might have expected that the additional spatial degree of freedom in 4D would trigger a transition to a new scaling regime at a comparatively lower $Ra$. However, our results show that the near-classical power law persists in 4D up to $Ra = 2 \times 10^8$. The robustness of these scaling exponents in 3D and 4D suggests that spatial dimensionality does not bypass the fundamental, classical mechanisms of heat transport. Consequently, what ultimately happens to the flow at exceptionally high Rayleigh numbers may remain essentially the same, demanding similarly high levels of thermal forcing for all spatial dimensions.

\section{Summary remarks and outlook}
\label{sec:summary}

In this study, we have explored the influence of spatial dimensionality, ranging from 1 to 4, on the dynamics of thermal convection. We find that the transient time required to achieve a statistically steady convective state is fundamentally tied to the geometric constraints of the system. In the degenerate 1D case, strict constraints prohibit convection physically and mathematically. In 2D, the transient times are exceptionally long and scale approximately linearly with the flow Reynolds number, posing a severe computational bottleneck for exploring very high-$Ra$ regimes. As the dimensionality increases to 3D and 4D, the dependence of the transient duration on $Re$ becomes markedly weaker. The temporal evolution of the kinetic energy in 4D closely mirrors that of 3D, confirming that the transition to a steady state becomes more efficient and less dependent on thermal forcing when sufficient spatial degrees of freedom are available.

Based on our findings on the transient time in systems of dimensions one to four, we construct the following empirical relation:
\begin{equation}
\mbox{Transient time} /t_f = \beta Re^\eta, \quad \mbox{where} \quad {\eta} = \frac{(D-3)(D-4)}{2(D-1)} \, .
\end{equation}
The prefactor $\beta \approx 40$ for 3D and 4D, whereas $\beta \approx 4.4 \times 10^{-3}$ for 2D.

We may now state our conclusions as follows. First, given the long transient times in
2D, it appears that a decisive study of the ultimate state of convection in 2D requires an unduly long computational time—which, in fact, may never be possible at extremely high Reynolds numbers. Second, for 3D convection, $t_{trns}$ is on the order of 50 free-fall times. Third, if it is possible to treat 3D convection as a perturbation of 4D convection, there may be some merit to studying 4D convection to determine various scaling problems, including the ultimate state. However, even if the 4D case behaves like the mean-field model and the 3D case can be conceived as a perturbation from 4D, we should expect that, at the level of accuracy with which we have been able to determine the transient times, we should not expect in this regard a huge differences between the 3D and 4D cases. 



\backsection[Acknowledgements]{
We thank Shreshthi for providing data of figure 3(b) in SM, and John Wettlaufer for comments on a preliminary draft. The authors gratefully acknowledge Dalma and Jubail clusters at NYU Abu Dhabi for providing computational resources. We also thank ALCF for computing time on Polaris via the Director's Discretionary Program.
}

\backsection[Funding]{This material is based upon work supported by Tamkeen under the NYU Abu Dhabi Research Institute grant G1502. A.P. also acknowledges financial support from ANRF (formerly SERB), India under the grant SRG/2023/001746. H.T. thanks IIT Kanpur for the Fellowship for Academic and Research Excellence (FARE). NYU supports the research of K.R.S.}

\backsection[Declaration of Interests]{The authors report no conflict of interest.}

\backsection[Data availability statement]{The data that support the findings of this study are available from the corresponding author upon reasonable request.}

\backsection[Author ORCIDs]{\\
A. Pandey, \href{https://orcid.org/0000-0001-8232-6626}{https://orcid.org/0000-0001-8232-6626};\\
H. Tiwari, \href{https://orcid.org/0009-0002-0116-9476}{https://orcid.org/0009-0002-0116-9476}; \\
K. R. Sreenivasan, \href{https://orcid.org/0000-0002-3943-6827}{https://orcid.org/0000-0002-3943-6827}.
}


\begin{thebibliography}{20}
\expandafter\ifx\csname natexlab\endcsname\relax\def\natexlab#1{#1}\fi
\def\au#1{#1} \def\ed#1{#1} \def\yr#1{#1}\def\at#1{#1}\def\jt#1{\textit{#1}} \def\bt#1{#1}\def\bvol#1{\textbf{#1}} \def\vol#1{#1} \def\pg#1{#1} \def\publ#1{#1}\def\arxiv#1{#1}\def\org#1{#1}\def\st#1{\textit{#1}}

\bibitem[{Chill\`{a}} \& {Schumacher}(2012)]{Chilla:EPJE2012}
{\sc \au{{Chill\`{a}}, F.} \& \au{{Schumacher}, J.}} \yr{2012}  \at{New perspectives in turbulent {R}ayleigh-{B}{\'e}nard convection}.  \jt{Eur. Phys. J. E}  \bvol{35},  \pg{58}.

\bibitem[{Fischer}(1997)]{Fischer:JCP1997}
{\sc \au{{Fischer}, P.~F.}} \yr{1997}  \at{An overlapping {S}chwarz method for spectral element solution of the incompressible {N}avier-{S}tokes equations}.  \jt{J. Comp. Phys.}  \bvol{133}~(1),  \pg{84--101}.

\bibitem[Goluskin {\em et~al.\/}(2014)Goluskin, Johnston, Flierl \& Spiegel]{Goluskin:JFM2014}
{\sc \au{Goluskin, D.}, \au{Johnston, H.}, \au{Flierl, G.~R.} \& \au{Spiegel, E.~A.}} \yr{2014}  \at{Convectively driven shear and decreased heat flux}.  \jt{J. Fluid Mech.}  \bvol{759},  \pg{360–385}.

\bibitem[Gotoh {\em et~al.\/}(2007)Gotoh, Watanabe, Shiga, Nakano \& Suzuki]{Gotoh:PRE2007}
{\sc \au{Gotoh, T.}, \au{Watanabe, Y.}, \au{Shiga, Y.}, \au{Nakano, T.} \& \au{Suzuki, E.}} \yr{2007}  \at{Statistical properties of four-dimensional turbulence}.  \jt{Phys. Rev. E}  \bvol{75},  \pg{016310}.

\bibitem[Horn {\em et~al.\/}(2013)Horn, Shishkina \& Wagner]{Horn:JFM2013}
{\sc \au{Horn, S.}, \au{Shishkina, O.} \& \au{Wagner, C.}} \yr{2013}  \at{On non-{O}berbeck-{B}oussinesq effects in three-dimensional {R}ayleigh-{B}{\'e}nard convection in glycerol}.  \jt{J. Fluid Mech.}  \bvol{724},  \pg{175–202}.

\bibitem[Kraichnan(1962)]{Kraichnan:POF1962}
{\sc \au{Kraichnan, R.~H.}} \yr{1962}  \at{Turbulent thermal convection at arbitrary {P}randtl number}.  \jt{Phys. Fluids}  \bvol{5}~(11),  \pg{1374--1389}.

\bibitem[Lindborg(2025)]{Lindborg:2025}
{\sc \au{Lindborg, E.}} \yr{2025} Scaling in two-dimensional {R}ayleigh-{B}\'enard convection,  \arxiv{arXiv: 2506.13213}.

\bibitem[Nelkin(1974)]{Nelkin:PRA1974}
{\sc \au{Nelkin, M.}} \yr{1974}  \at{Turbulence, critical fluctuations, and intermittency}.  \jt{Phys. Rev. A}  \bvol{9},  \pg{388--395}.

\bibitem[Nelkin(1975)]{Nelkin:PRA1975}
{\sc \au{Nelkin, M.}} \yr{1975}  \at{Scaling theory of hydrodynamic turbulence}.  \jt{Phys. Rev. A}  \bvol{11},  \pg{1737--1743}.

\bibitem[Pandey(2021)]{Pandey:JFM2021}
{\sc \au{Pandey, A.}} \yr{2021}  \at{Thermal boundary layer structure in low-{P}randtl-number turbulent convection}.  \jt{J. Fluid Mech.}  \bvol{910},  \pg{A13}.

\bibitem[Pandey {\em et~al.\/}(2026)Pandey, Schumacher, Parsani \& Sreenivasan]{Pandey:JFM2026}
{\sc \au{Pandey, A.}, \au{Schumacher, J.}, \au{Parsani, M.} \& \au{Sreenivasan, K.~R.}} \yr{2026}  \at{Influence of plume activity on thermal convection in a rectangular cell}.  \jt{J. Fluid Mech.}  \bvol{1034},  \pg{A41}.

\bibitem[Pandey \& Sreenivasan(2021)]{Pandey:EPL2021}
{\sc \au{Pandey, A.} \& \au{Sreenivasan, K.~R.}} \yr{2021}  \at{Convective heat transport in slender cells is close to that in wider cells at high {R}ayleigh and {P}randtl numbers}.  \jt{Europhys. Lett.}  \bvol{135}~(2),  \pg{24001}.

\bibitem[Pandey \& Sreenivasan(2025)]{Pandey:JFM2025}
{\sc \au{Pandey, A.} \& \au{Sreenivasan, K.~R.}} \yr{2025}  \at{Transient and steady convection in two dimensions}.  \jt{J. Fluid Mech.}  \bvol{1015},  \pg{A42}.

\bibitem[Pandey \& Verma(2016)]{Pandey:PoF2016}
{\sc \au{Pandey, A.} \& \au{Verma, M.~K.}} \yr{2016}  \at{Scaling of large-scale quantities in {R}ayleigh-{B}{\'e}nard convection}.  \jt{Phys. Fluids}  \bvol{28}~(9),  \pg{095105}.

\bibitem[Pandey {\em et~al.\/}(2014)Pandey, Verma \& Mishra]{Pandey:PRE2014}
{\sc \au{Pandey, A.}, \au{Verma, M.~K.} \& \au{Mishra, P.~K.}} \yr{2014}  \at{Scaling of heat flux and energy spectrum for very large {P}randtl number convection}.  \jt{Phys. Rev. E}  \bvol{89},  \pg{023006}.

\bibitem[Scheel \& Schumacher(2017)]{Scheel:PRF2017}
{\sc \au{Scheel, J.~D.} \& \au{Schumacher, J.}} \yr{2017}  \at{Predicting transition ranges to fully turbulent viscous boundary layers in low {P}randtl number convection flows}.  \jt{Phys. Rev. Fluids}  \bvol{2},  \pg{123501}.

\bibitem[{Silano} {\em et~al.\/}(2010){Silano}, {Sreenivasan} \& {Verzicco}]{Silano:JFM2010}
{\sc \au{{Silano}, G.}, \au{{Sreenivasan}, K.~R.} \& \au{{Verzicco}, R.}} \yr{2010}  \at{Numerical simulations of {R}ayleigh-{B}{\'e}nard convection for {P}randtl numbers between $10^{-1}$ and $10^4$ and {R}ayleigh numbers between $10^5$ and $10^9$}.  \jt{J. Fluid Mech.}  \bvol{662},  \pg{409--446}.

\bibitem[{Stevens} {\em et~al.\/}(2010){Stevens}, {Verzicco} \& {Lohse}]{Stevens:JFM2010}
{\sc \au{{Stevens}, R.}, \au{{Verzicco}, R.} \& \au{{Lohse}, D.}} \yr{2010}  \at{Radial boundary layer structure and {N}usselt number in {R}ayleigh-{B}{\'e}nard convection}.  \jt{J. Fluid Mech.}  \bvol{643},  \pg{495--507}.

\bibitem[Wilson \& Fisher(1972)]{Wilson:PRL1972}
{\sc \au{Wilson, K.~G.} \& \au{Fisher, M.~E.}} \yr{1972}  \at{Critical exponents in 3.99 dimensions}.  \jt{Phys. Rev. Lett.}  \bvol{28},  \pg{240--243}.

\bibitem[Yamamoto {\em et~al.\/}(2012)Yamamoto, Shimizu, Inoshita, Nakano \& Gotoh]{Yamamoto:PRE2012}
{\sc \au{Yamamoto, T.}, \au{Shimizu, H.}, \au{Inoshita, T.}, \au{Nakano, T.} \& \au{Gotoh, T.}} \yr{2012}  \at{Local flow structure of turbulence in three, four, and five dimensions}.  \jt{Phys. Rev. E}  \bvol{86},  \pg{046320}.

\end{thebibliography}

\clearpage
\includepdf[pages=-,fitpaper=true]{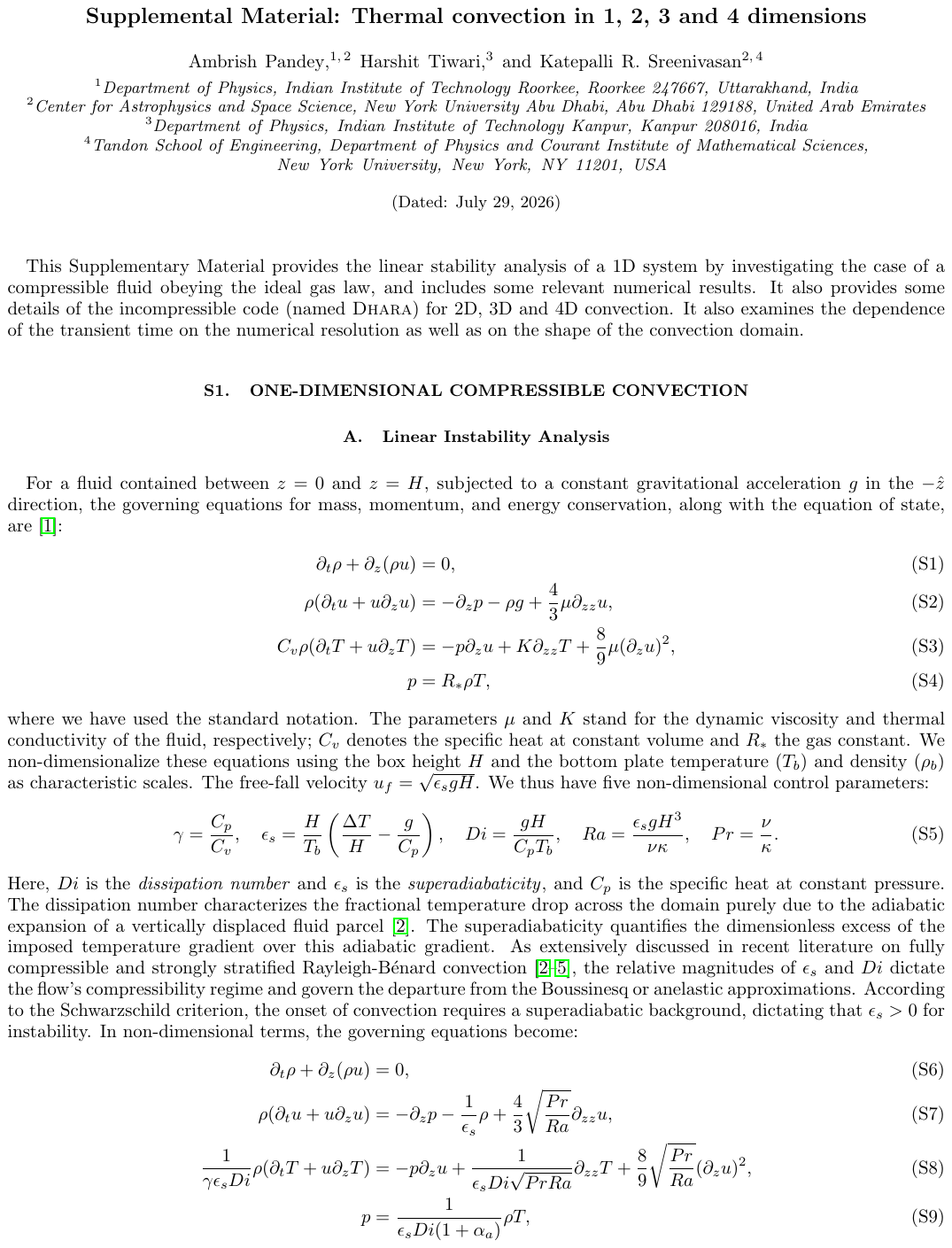}


\end{document}